\documentclass[global]{svjour}
\usepackage{amssymb}
\usepackage{amsfonts}
\usepackage{amsmath}

\input{tcilatex}
\begin{document}

\journalname{EPJC}
\title{Quantization of rotating linear dilaton black holes}
\author{I. Sakalli\inst{1}}
\institute{Department of Physics, Eastern Mediterranean University, Gazimagosa, North
Cyprus, Mersin 10, Turkey. 
\email{izzet.sakalli@emu.edu.tr}
}
\dedication{}
\offprints{}
\mail{}
\maketitle

\begin{abstract}
In this paper, we focus on the quantization of $4$--dimensional rotating
linear dilaton black hole (RLDBH) spacetime describing an action, which
emerges in the Einstein--Maxwell-Dilaton-Axion (EMDA) theory. RLDBH
spacetime has a non-asymptotically flat (NAF) geometry. When the rotation
parameter \textquotedblleft $a$\textquotedblright\ vanishes, the spacetime
reduces to its static form, the so-called linear dilaton black hole (LDBH)
metric. Under scalar perturbations, we show that the radial equation reduces
to a hypergeometric differential equation. Using the boundary conditions of
the quasinormal modes (QNMs), we compute the associated complex frequencies
of the QNMs. In a particular case, QNMs are applied in the rotational
adiabatic invariant quantity, and we obtain the quantum entropy/area spectra
of the RLDBH. Both spectra are found to be discrete and equidistant, and
independent of $a-$parameter despite the modulation of QNMs by this
parameter.
\end{abstract}


\section{Introduction}

Quantization of BHs has always interested physicists working on quantum
gravity theory. The topic was introduced in the 1970s by Bekenstein, who
proved that the entropy of a BH is proportional to the area of the BH
horizon \cite{Bek1,Bek2}. The equidistant area spectrum of a BH \cite%
{Bek3,Bek4,Bek5} is given by

\begin{equation}
A_{n}^{BH}=8\pi \xi \hbar =\epsilon nl_{p}^{2},\text{ \ \ \ \ \ \ }%
(n=0,1,2.......),  \label{1}
\end{equation}

where $A_{n}^{BH}$\ denotes the area spectrum of the BH horizon, $n$ is the
associated quantum number, $\xi $ represents a number that is order of
unity, $\epsilon $ is a dimensionless constant, and $l_{p}$ is the Planck
length. In his celebrated studies, Bekenstein considered the BH horizon area
as an adiabatic invariant quantity, and derived the discrete equally spaced
area spectrum by Ehrenfest's principle. Eq. (1) shows that when a test
particle is swallowed by a BH, the horizon area increases by a minimum of $%
\Delta A_{\min }^{BH}=\epsilon l_{p}^{2}$. Throughout this paper, we
normalize the fundamental units as $c=G=1$ and $l_{p}^{2}=\hbar $. According
to Bekenstein \cite{Bek3}, a BH\ horizon is made by patches of equal area $%
\epsilon \hbar ,$ where $\epsilon =8\pi $. However, although the derivation
of evenly spaced area spectra has received much attention, the value of $%
\epsilon $ has been somewhat controversial, because it depends on the method
used to obtain the spectrum (for a topical review, the reader is referred to 
\cite{Gua} and references therein).

QNMs, known as the characteristic ringing of BHs, are damped oscillations
characterized by a discrete set of complex frequencies. They are determined
by solving the wave equation under perturbations of the BH by an external
field. A perturbed BH tends to equilibrate itself by emitting energy in the
form of gravitational waves. For this reason, QNMs are also important for
experiments such as LIGO \cite{Ligo}, which aims to detect gravitational
wave phenomena. Inspired by the above findings, Hod \cite{Hod1,Hod2}
hypothesized that $\epsilon $ can be computed from the QNMs of a vibrating
BH. To this end, Hod adopted Bohr's correspondence principle \cite{Bohr},
and conjectured that the real part of the asymptotic QNM frequency ($\func{Re%
}\omega $) of a BH relates to the quantum transition energy between
sequential quantum levels of the BH. Thus, this vibrational frequency
induces a change $\Delta M=\hbar \left( \func{Re}\omega \right) $ in the BH
mass. For a Schwarzschild BH, Hod's calculations yield $\epsilon =4\ln 3$ 
\cite{Hod1}. Later, Kunstatter \cite{Kunstatter} used the natural adiabatic
invariant quantity $I_{adb}$, which is defined for a system with energy $E$
as

\begin{equation}
I_{adb}=\int \frac{dE}{\Delta \omega },  \label{2}
\end{equation}

where $\Delta \omega =\omega _{n-1}-\omega _{n}$ is the transition
frequency. At large quantum numbers ($n\rightarrow \infty $), the
Bohr--Sommerfeld quantization condition applies and $I_{adb}$ behaves as a
quantized quantity ($I_{adb}\simeq n\hbar $). By calculating $\Delta \omega $%
, using the real part $\func{Re}\omega $, and replacing $E$\ with the mass $%
M $ of the static BH in Eq. (2), Kunstatter retrieved Hod's result $\epsilon
=4\ln 3$ for a Schwarzschild BH. In 2008, Maggiore \cite{Maggiore} proposed
that the proper physical frequency of the harmonic oscillator with a damping
term takes the form $\omega =\sqrt{\func{Re}\omega ^{2}+\func{Im}\omega ^{2}}
$, where $\func{Im}\omega $ is the imaginary part of the QNM frequency.
Actually, this form of the proper physical frequency was first proposed for
QNM frequencies by Wang et al. \cite{BWang}. Thus, Hod's result \cite{Hod1}
is obtained whenever $\func{Re}\omega \gg \func{Im}\omega $.

On the other hand, as is well-known, highly damped (excited) QNMs correspond
to $\func{Im}\omega \gg \func{Re}\omega $. Consequently $\Delta \omega
\approx \func{Im}\omega _{n-1}-\func{Im}\omega _{n}$, which is directly
proportional to the Hawking temperature ($T_{H}$) of the BH \cite{Nollert}.
This new interpretation recovers the original result of Bekenstein; i.e., $%
\epsilon =8\pi ,$ for a Schwarzschild BH. Since its inception, Maggiore's
method (MM) has been widely studied by other researchers investigating
various BH backgrounds (see for example \cite%
{Samp1,Samp2,Samp3,Samp4,Samp6,Samp7,Samp8,Samp9,Samp10,Samp11}).

Thereafter, Vagenas \cite{Vagenas} and Medved \cite{Medved} amalgamated the
results of Kunstatter and Maggiore, and proposed the following rotational
version of the adiabatic invariant:

\begin{equation}
I_{adb}^{rot}=\int \frac{dM-\Omega _{H}dJ}{\Delta \omega }=n\hbar ,
\label{3n}
\end{equation}

where $\Omega _{H}$ and $J$ represent angular velocity and angular momentum
of the rotating BH, respectively. Vagenas and Medved \cite{Vagenas,Medved}
inserted the asymptotic form of the QNMs of the Kerr BH \cite{Hod3} into the
above expression and obtained an equidistant area spectrum conditional on $%
M^{2}\gg J$.

Following Refs. \cite{ChenYang,MajhiBest,WenWei,MajhiPLB,MajhiPLB2,IOPconf},
one can use the first law of thermodynamics ($T_{H}dS^{BH}=dM-\Omega _{H}dJ$%
) to safely derive the entropy spectra from the QNMs via

\begin{equation}
I_{adb}^{rot}=\int \frac{T_{H}dS^{BH}}{\Delta \omega }=n\hbar .  \label{4n}
\end{equation}

In the present study, we analyze the entropy/area spectra of the RLDBHs. The
RLDBHs, developed by Cl\'{e}ment, Gal'tsov, and Leygnac \cite{Clement}, are
the NAF solutions to the EMDA theory in four dimensions. These BHs have both
dilaton and axion fields. Especially, the axion field may legitimate the
existence of the cold dark matter \cite{Axion,Axion2} in the RLDBH
spacetime. Although many studies have focused on non--rotating forms of
these BHs (known as LDBHs) and their related topics such as Hawking
radiation, entropic forces, higher dimensions, quantization, and the
gravitational lensing effect on the Hawking temperature \cite%
{Fabris,Iz1,Iz2,Iz3,Iz4,Iz5,Iz6,Iz7,Iz8,Iz9,Slavov}, studies on RLDBHs have
remained very limited. To our knowledge, RLDBH studies have focused on
branes, holography, greybody factors, Hawking radiation, and QNMs \cite%
{rldbh1,rldbh2,rldbh3}, while ignoring the quantization of the RLDBH. Our
present study bridges this gap in the literature.

We compute the QNMs of the RLDBH by giving an analytical exact solution (as
being an alternative solution to the solution presented by Li \cite{rldbh3})
to the massless Klein--Gordon equation (KGE) in the RLDBH geometry. To this
end, we use a particular transformation for the hypergeometric function, and
impose the appropriate boundary conditions for the QNMs, which obey only
ingoing wave at the event horizon and only outgoing wave at spatial
infinity. Our results support the study of Li \cite{rldbh3}. On the other
hand, we remark that equation (45) of \cite{rldbh3} provides only one of the
two sets of the QNMs of the RLDBHs and is independent of the rotation
parameter $a$. However, this parameter is required in a general form of the
QNMs of the RLDBH. In fact, by inserting $\mathbf{b}_{s}=-n$ instead of $%
\mathbf{a}_{s}=-n$ (see equations (17) and (44) of \cite{rldbh3}) in the
QNMs evaluation, we derive a further set of QNMs involving the rotation term 
$a$, which reduces to the QNMs of the LDBHs as $a\rightarrow 0$. Somehow
this possibility has been overlooked in \cite{rldbh3}. In the present paper,
we show that, in a particular case, the QNMs with rotation parameter $a$ can
be represented as a simple expression including the angular velocity and
surface gravity terms, as previously obtained for the Kerr BH \cite{Hod3}.
We then derive the equally spaced entropy/area spectra of the RLDBHs.

The remainder of this paper is arranged as follows. Sec. 2 introduces the
RLDBH metric, and analyzes the KGE for a massless scalar field in this
geometry. In Sec. 3, we show that the KGE reduces to a hypergeometric
differential equation. We also read the QNMs of the RLDBHs and use one of
their two sets to derive the entropy/area spectra of the RLDBHs. Conclusions
are presented in Sec. 4.

\section{RLDBH Spacetime and Separation of KGE}

In this section, we provide a general overview of the RLDBH solution and
present its basic thermodynamic features. We then investigate the KGE in the
geometry of the RLDBH. We show that after separation by variables, the
radial equation can be obtained.

The action of the EMDA theory, which comprises the dilaton field $\phi $ and
the axion (pseudoscalar) $\varkappa $ coupled to an Abelian vector field $%
\mathcal{A}$ is given by

\begin{eqnarray}
\mathit{S} &=&\frac{1}{16\pi }\int d^{4}x\sqrt{\left\vert g\right\vert }%
\left\{ -R+2\partial _{\mu }\phi \partial ^{\mu }\phi +\frac{1}{2}e^{4\phi
}\partial _{\mu }\varkappa \partial ^{\mu }\varkappa -\right.  \notag \\
&&\left. e^{-2\phi }F_{\mu \nu }F^{\mu \nu }-\varkappa F_{\mu \nu }%
\widetilde{F}^{\mu \nu }\right\} ,  \label{5n}
\end{eqnarray}

where $F_{\mu \upsilon }$ is the Maxwell two--form associated with a $U(1)$
subgroup of $E_{8}\times E_{8}$ or Spin(32)/$Z_{2}$, and $\widetilde{F}^{\mu
\nu }$ denotes the dual of $F_{\mu \upsilon }$. In the absence of a NUT
charge, the RLDBH solution is given by \cite{Clement}

\begin{equation}
ds^{2}=-f(r)dt^{2}+\frac{dr^{2}}{f(r)}+R(r)\left[ d\theta ^{2}+\sin
^{2}\theta \left( d\varphi -\frac{a}{R(r)}dt\right) ^{2}\right] ,  \label{6n}
\end{equation}

where the metric functions are given by%
\begin{equation}
R(r)=rr_{0},  \label{7n}
\end{equation}

\begin{equation}
f(r)=\frac{\Lambda }{R(r)},  \label{8n}
\end{equation}

In Eq. (8) $\Lambda =(r-r_{2})(r-r_{1})$, where $r_{1}$ and $r_{2}$ denote
the inner and outer\ (event) horizons, respectively, under the condition $%
f(r)=0$. These radii are given by

\begin{equation}
r_{j}=M+(-1)^{j}\sqrt{M^{2}-a^{2}},\text{ \ \ \ \ \ (}j=1,2\text{),}
\label{9n}
\end{equation}

where the integration constant $M$ is related to the quasilocal mass ($%
M_{QL} $) of the RLDBH, and $a$ is the abovementioned rotation parameter
determining the angular momentum of the RLDBH. The background electric
charge $Q$ is related to the constant parameter $r_{0}$ by $\sqrt{2}Q=r_{0}$%
. Furthermore, the background fields are given by

\begin{equation}
e^{-2\phi }=\frac{R(r)}{r^{2}+a^{2}\cos ^{2}\theta },  \label{10n}
\end{equation}

\begin{equation}
\varkappa =-\frac{r_{0}a\cos \theta }{r^{2}+a^{2}\cos ^{2}\theta },
\label{11n}
\end{equation}

The Maxwell field $\left( F=d\mathcal{A}\right) $ is derivable from the
following electromagnetic four--vector potential

\begin{equation}
\mathcal{A=}\frac{1}{\sqrt{2}}(e^{2\phi }dt+a\sin ^{2}\theta d\varphi ).
\label{12n}
\end{equation}

From Eq. (9), we infer that for a BH

\begin{equation}
M\geq a.  \label{13n}
\end{equation}

Meanwhile, because the RLDBH\ has a NAF\ geometry, $M_{QL}$ can be defined
by the formalism of Brown and York\ \cite{BrownYork}. Thus, we obtain

\begin{equation}
M_{QL}=\frac{M}{2}.  \label{14n}
\end{equation}

The angular momentum $J$ is computed as

\begin{equation}
J=\frac{ar_{0}}{2}.  \label{15n}
\end{equation}

According to the definition given by Wald \cite{Wald}, the surface gravity
of the RLDBH is simply evaluated as follows:

\begin{equation}
\kappa =\left. \frac{f^{\prime }(r)}{2}\right\vert _{r=r_{2}}=\frac{\left(
r_{2}-r_{1}\right) }{2r_{2}r_{0}},  \label{16n}
\end{equation}

where a prime on a function denotes differentiation with respect to its
argument. Thus, the Hawking temperature \cite{Wald,Rovelli} of the RLDBH is
obtained as

\begin{align}
T_{H}& =\frac{\hbar \kappa }{2\pi },  \notag \\
& =\frac{\hbar \left( r_{2}-r_{1}\right) }{4\pi r_{2}r_{0}}.  \label{17n}
\end{align}

Clearly, the temperature vanishes in the extreme case ($M=a$) since $%
r_{2}=r_{1}$. The Bekenstein-Hawking entropy of a BH is defined in
Einstein's theory as \cite{Wald}

\begin{equation}
S^{BH}=\frac{A^{BH}}{4\hbar }=\frac{\pi r_{2}r_{0}}{\hbar },  \label{18n}
\end{equation}

where $A^{BH}$ is area of the surface of the event\ horizon. To satisfy the
first law of BH\ mechanics, we also need the angular velocity which takes
the following form for the RLDBH:

\begin{equation}
\Omega _{H}=\frac{a}{r_{2}r_{0}}.  \label{19n}
\end{equation}

We can now validate the first law of thermodynamics in the RLDBH through the
relation

\begin{equation}
dM_{QL}=T_{H}dS^{BH}+\Omega _{H}dJ.  \label{20n}
\end{equation}

It is worth mentioning that Eq. (20) is independent of electric charge $Q$.
Here, $Q$ is considered as a background charge of fixed value -- a
characteristic feature of linear dilaton backgrounds \cite{Clement}.

To obtain the RLDBH entropy spectrum via the MM, we first consider the
massless KGE on that geometry. The general massless KGE in a curved
spacetime is given by

\begin{equation}
\partial _{j}(\sqrt{-g}\partial ^{j}\Phi )=0,\text{ \ \ \ }j=0...3.
\label{21n}
\end{equation}

We invoke the following ansatz for the scalar field $\Phi $ in the above
equation

\begin{equation}
\Phi =\frac{\Re (r)}{\sqrt{r}}e^{-i\omega t}Y_{l}^{m}(\theta ,\varphi ),%
\text{ \ \ }Re(\omega )>0,  \label{22n}
\end{equation}

where $Y_{l}^{m}(\theta ,\varphi )$ denote the well-known spheroidal
harmonics with eigenvalues $\lambda =-l(l+1)$ \cite{Du}. Here, $m$ and $l$
denote the magnetic quantum number and orbital angular quantum number,
respectively. Thus, we obtain the following radial equation

\begin{equation}
\Lambda \Re ^{\prime \prime }(r)+(\Lambda ^{\prime }-\frac{\Lambda }{r})\Re
^{\prime }(r)+\left[ \frac{\left( \omega rr_{0}-ma\right) ^{2}}{\Lambda }%
+\lambda +\frac{3\Lambda }{4r^{2}}-\frac{\Lambda ^{\prime }}{2r}\right] \Re
(r)=0.  \label{23n}
\end{equation}

The tortoise coordinate $r^{\ast }$ is defined as \cite{Chandra}

\begin{equation}
r^{\ast }=\int \frac{dr}{f(r)},  \label{24n}
\end{equation}

whereby

\begin{equation}
r^{\ast }=\frac{r_{0}}{r_{2}-r_{1}}\ln \left[ \frac{(\frac{r}{r_{2}}%
-1)^{r_{2}}}{(r-r_{1})^{r_{1}}}\right] .  \label{25n}
\end{equation}

It is easily checked that the asymptotic limits of $r^{\ast }$ are

\begin{equation}
\lim_{r\rightarrow r_{2}}r^{\ast }=-\infty \text{ \ \ and \ \ }%
\lim_{r\rightarrow \infty }r^{\ast }=\infty .  \label{26n}
\end{equation}

\section{QNMs and entropy/area spectra of RLDBHs}

In this section, we obtain an exact analytical solution to the radial
equation (23). In computing the QNM frequencies, we impose the relevant
boundary conditions: (i) the field is purely ingoing at the horizon; (ii)
the field is purely outgoing at spatial infinity. Finally, we represent how
the quantum spectra of the RLDBH can be equidistant and fully agree with the
Bekenstein's conjecture (1).

In order to solve Eq. (23) analytically, one can redefine the $r$%
--coordinate with a new variable $x$:

\begin{equation}
x=\frac{r-r_{2}}{r_{2}-r_{1}}.  \label{27n}
\end{equation}

Thus, the radial equation (23) adopts the form

\begin{equation}
x(1+x)\Re ^{\prime \prime }(x)+(1+2x)\Re ^{\prime }(x)+\Theta \Re (x)=0,
\label{28n}
\end{equation}

where

\begin{equation}
\Theta =\frac{\left\{ r_{0}\omega \left[ x\left( r_{2}-r_{1}\right) +r_{2}%
\right] -ma\right\} ^{2}}{\left( r_{2}-r_{1}\right) ^{2}x\left( 1+x\right) }%
+\lambda .  \label{29n}
\end{equation}

The above equation can be rewritten as

\begin{equation}
\Theta =\frac{A^{2}}{x}+\frac{B^{2}}{1+x}-C,  \label{30n}
\end{equation}

where

\begin{equation}
A=\frac{\widehat{\omega }}{2\kappa },  \label{31n}
\end{equation}

\begin{equation}
B=i\frac{\omega r_{1}-m\Omega _{H}r_{2}}{2\kappa r_{2}},  \label{32n}
\end{equation}

\begin{equation}
C=-\left( \lambda +\omega ^{2}r_{0}^{2}\right) ,  \label{33n}
\end{equation}

and

\begin{equation}
\widehat{\omega }=\omega -m\Omega _{H}.  \label{34n}
\end{equation}

The solution of the radial equation (28) reads as follows (see Appendix A):

\begin{equation}
\Re (x)=C_{1}(x)^{iA}(1+x)^{-B}F\left( a_{n},b_{n};c_{n};-x\right)
+C_{2}(x)^{-iA}(1+x)^{-B}F\left( \widetilde{a}_{n},\widetilde{b}_{n};%
\widetilde{c}_{n};-x\right) ,  \label{35n}
\end{equation}

where $F\left( a_{n},b_{n};c_{n};-x\right) $ is the hypergeometric function 
\cite{Abram} and $C_{1},$ $C_{2}$ are constants. Here, the coefficients $%
\widetilde{a}_{n}$, $\widetilde{b}_{n}$, and $\widetilde{c}_{n}$ are given
through the relations

\begin{eqnarray}
\widetilde{a}_{n} &=&a_{n}-c_{n}+1,  \notag \\
\widetilde{b}_{n} &=&b_{n}-c_{n}+1,  \notag \\
\widetilde{c}_{n} &=&2-c_{n},  \label{36n}
\end{eqnarray}

where 
\begin{equation}
a_{n}=\frac{1}{2}-B+iA+\frac{\sqrt{4C+1}}{2},  \label{37n}
\end{equation}

\begin{equation}
b_{n}=\frac{1}{2}-B+iA-\frac{\sqrt{4C+1}}{2},  \label{38n}
\end{equation}

\begin{equation}
c_{n}=1+2iA.  \label{39n}
\end{equation}

One can see that

\begin{equation}
-B+iA=i\omega r_{0},  \label{40n}
\end{equation}

\begin{equation}
\frac{\sqrt{4C+1}}{2}=\eta =ir_{0}\sqrt{\omega ^{2}-\tau ^{2}},  \label{41n}
\end{equation}

where

\begin{equation}
\tau =\frac{l+1/2}{r_{0}}.  \label{42n}
\end{equation}

In the neighborhood of the horizon ($x\rightarrow 0$), the function $\Re (x)$
behaves as

\begin{equation}
\Re (x)=C_{1}(x)^{iA}+C_{2}(x)^{-iA}.  \label{43n}
\end{equation}

If we expand the metric function $f(r)$ as a Taylor series with respect to $%
r $ around the event horizon $r_{2}$:

\begin{align}
f(r)& \simeq f^{\prime }(r_{2})(r-r_{2})+\Game (r-r_{2})^{2},  \notag \\
& \simeq 2\kappa (r-r_{2}),  \label{44n}
\end{align}

the tortoise coordinate (24) then transforms into

\begin{equation}
r^{\ast }\simeq \frac{1}{2\kappa }\ln x.  \label{45n}
\end{equation}

Therefore, the scalar field near the horizon can be written in the following
way:

\begin{equation}
\Phi \sim C_{1}e^{-i(\omega t-\widehat{\omega }r^{\ast })}+C_{2}e^{-i(\omega
t+\widehat{\omega }r^{\ast })}.  \label{46n}
\end{equation}

From the above expression, we see that the first term corresponds to an
outgoing wave, while the second one represents an ingoing wave \cite%
{rldbh3,Starobinskii}. For computing the QNMs, we have to impose that there
exist only ingoing waves at the horizon so that, in order to satisfy the
condition, we set $C_{1}=0$. Then the physically acceptable radial solution
(35) is given by

\begin{equation}
\Re (x)=C_{2}(x)^{-iA}(1+x)^{-B}F\left( \widetilde{a}_{n},\widetilde{b}_{n};%
\widetilde{c}_{n};-x\right) .  \label{47n}
\end{equation}

For matching the near--horizon and far regions, we are interested in the
large $r$ behavior ($r\simeq r^{\ast }\simeq x\simeq 1+x\rightarrow \infty $%
) of the solution (47). To achieve this aim, one uses $z\rightarrow \frac{1}{%
z}$ transformation law for the hypergeometric function \cite{Abram,Olver}

\begin{align}
F(\widetilde{a},\widetilde{b};\widetilde{c};z)& =\frac{\Gamma (\widetilde{c}%
)\Gamma (\widetilde{b}-\widetilde{a})}{\Gamma (\widetilde{b})\Gamma (%
\widetilde{c}-\widetilde{a})}(-z)^{-\widetilde{a}}F(\widetilde{a},\widetilde{%
a}+1-\widetilde{c};\widetilde{a}+1-\widetilde{b};1/z)  \notag \\
& +\frac{\Gamma (\widetilde{c})\Gamma (\widetilde{a}-\widetilde{b})}{\Gamma (%
\widetilde{a})\Gamma (\widetilde{c}-\widetilde{b})}(-z)^{-\widetilde{b}}F(%
\widetilde{b},\widetilde{b}+1-\widetilde{c};\widetilde{b}+1-\widetilde{a}%
;1/z).  \label{48n}
\end{align}

Thus, we obtain the asymptotic behavior of the radial solution (47) as

\begin{equation}
\Re (r)=\frac{C_{2}}{\sqrt{r}}\left[ \frac{\Gamma (\widetilde{c}_{n})\Gamma (%
\widetilde{b}_{n}-\widetilde{a}_{n})}{\Gamma (\widetilde{b}_{n})\Gamma (%
\widetilde{c}_{n}-\widetilde{a}_{n})}\left( r\right) ^{-\eta }+\frac{\Gamma (%
\widetilde{c})\Gamma (\widetilde{a}-\widetilde{b})}{\Gamma (\widetilde{a}%
)\Gamma (\widetilde{c}-\widetilde{b})}\left( r\right) ^{\eta }\right] .
\label{49n}
\end{equation}

Since the QNMs impose that the ingoing waves spontaneously terminate at
spatial infinity (meaning that only purely outgoing waves are allowed), the
first term of Eq. (49) vanishes. This scenario is enabled by the poles of
the Gamma function in the denominator of the second term; $\Gamma (%
\widetilde{c}_{n}-\widetilde{a}_{n})$ or $\Gamma (\widetilde{b}_{n})$. Since
the Gamma function $\Gamma (\sigma )$ has the poles at $\sigma =-n$ for $%
n=0,1,2,...$, the wave function satisfies the considered boundary condition
only upon the following restrictions:

\begin{equation}
\widetilde{c}_{n}-\widetilde{a}_{n}=1-a_{n}=-n,  \label{50n}
\end{equation}

or

\begin{equation}
\widetilde{b}_{n}=-n.  \label{51n}
\end{equation}

These conditions determine the QNMs. From Eqs. (37) and (50), we find

\begin{eqnarray}
\omega _{n} &=&-\frac{i}{(2n+1)r_{0}}\left[ n(n+1)-l(l+1)\right] ,  \notag \\
&=&-\frac{i}{r_{0}}\left[ n+\frac{1}{2}-\frac{\left( 2l+1\right) ^{2}}{%
2(2n+1)}\right] ,  \label{52n}
\end{eqnarray}

which is nothing but the QNM solution represented in \cite{rldbh3}.
Obviously, it is independent from the rotation parameter $a$ of the RLDBH.
For the highly damped modes ($n\rightarrow \infty $), it reduces to

\begin{equation}
\omega _{n}=-i(n+\frac{1}{2})\kappa _{0},\text{ \ \ \ \ \ \ (}n\rightarrow
\infty \text{),}  \label{53n}
\end{equation}

where $\kappa _{0}=\frac{1}{2r_{0}}$\ is the surface gravity of the LDBH
(i.e., $a=0$). Hence, its corresponding transition frequency becomes

\begin{eqnarray}
\Delta \omega &\approx &\func{Im}\omega _{n-1}-\func{Im}\omega _{n},  \notag
\\
&=&\kappa _{0}.  \label{54n}
\end{eqnarray}

QNMs obtaining from Eq. (51) have a rather complicated form. However, if we
consider the LDBH case, simply setting $a=0$ in Eq. (51), one promptly reads
the QNMs as represented in Eq. (52). On the other hand, for $a\neq 0$ case,
we inferred from our detailed analysis that the quantization of the RLDBH,
which is fully consistent with the Bekenstein's conjecture (1) can be
achieved when the case of large charge\ ($r_{0}\gg $) and low orbital
quantum number $l$ is considered. Hence $\tau \ll $ (see Eq. (42)), and
therefore $\eta \approx i\omega r_{0}$ in Eq. (41). In this case, Eq. (51)
becomes 
\begin{equation}
\widetilde{b}_{n}\approx \frac{1}{2}-i\frac{\widehat{\omega }}{2\kappa }=-n.
\label{55n}
\end{equation}

Thus, we compute the QNMs as follows:

\begin{equation}
\omega _{n}=m\Omega _{H}-i(n+\frac{1}{2})\kappa .  \label{56n}
\end{equation}

The structural form of the above result corresponds to the Hod's QNM result
obtained for the Kerr BH \cite{Hod3}. Now, the transition frequency between
two highly damped neighboring states is

\begin{eqnarray}
\Delta \omega &\approx &\func{Im}\omega _{n-1}-\func{Im}\omega _{n},\text{ \
\ \ \ \ \ (}n\rightarrow \infty \text{)}  \notag \\
&=&\kappa ,  \notag \\
&=&2\pi T_{H}/\hbar ,  \label{57n}
\end{eqnarray}

which reduces to Eq. (54) when setting $a=0$ ($\kappa \rightarrow \kappa
_{0} $). Thus, after substituting Eq. (57) into Eq. (4), the adiabatic
invariant quantity reads

\begin{equation}
I_{adb}^{rot}=\frac{\hbar S^{BH}}{2\pi }=\frac{r_{2}r_{0}}{2}=n\hbar .
\label{58n}
\end{equation}

From Eq. (58), the entropy spectrum is obtained as

\begin{equation}
S_{n}^{BH}=2\pi n.  \label{59n}
\end{equation}

Because $S^{BH}=\frac{A^{BH}}{4\hbar },$ the area spectrum is then obtained
as

\begin{equation}
A_{n}^{BH}=8\pi n\hbar ,  \label{60n}
\end{equation}

and the minimum area spacing becomes

\begin{equation}
\Delta A_{\min }^{BH}=8\pi \hbar .  \label{61n}
\end{equation}

The above results imply that the spectral-spacing coefficient becomes $%
\epsilon =8\pi $, which is fully in agreement with the Bekenstein's original
result \cite{Bek3,Bek4,Bek5}. Furthermore, the spectra of the RLDBH are
clearly independent of the rotation parameter $a$.

\section{Conclusion}

In this paper, we quantized the entropy and horizon area of the RLDBH.
According to the MM, which considers the perturbed BH as a highly damped
harmonic oscillator, the transition frequency in Eq. (4) (describing $%
I_{adb}^{rot}$) is governed by $\func{Im}\omega $ rather than by $\func{Re}%
\omega $ of the QNMs. To compute $\func{Im}\omega $, we solved the radial
equation (23) as a hypergeometric differential equation. The parameters of
the hypergeometric functions were rigorously found by straightforward
calculation. We then computed the complex QNM frequencies of the RLDBH. In
the case of $\tau \ll $, we showed that the obtained QNMs with the rotation
term $a$ are in the form of Hod's asymptotic QNM formula \cite{Hod3}. After
obtaining the transition frequency of the highly damped QNMs, we derived the
quantized entropy and area spectra of the RLDBH. Both spectra are equally
spaced and independent of the rotation parameter $a$. Our results support
the Kothawala et al.'s conjecture \cite{Kothawala}, which states that the
area spectrum is equally spaced under Einstein's gravity theory but
unequally spaced under alternative (higher-derivative) gravity theories. On
the other hand, the numerical coefficient was evaluated as $\epsilon =8\pi $
or $\xi =1$, which is consistent with previous results \cite%
{Bek3,Bek4,Bek5,Maggiore}.

In future work, we will extend our analysis to the Dirac equation, which can
be formulated in the Newman--Penrose picture \cite{Chandra,SDirac1,SDirac2}.
In this way, we plan to analyze the quantization of stationary spacetimes
using fermion QNMs.

\begin{quote}
{\LARGE Appendix A}
\end{quote}

By redefining the independent variable $x=-y$, Eq. (28) becomes

\begin{equation}
y(1-y)\Re ^{\prime \prime }(y)+(1-2y)\Re ^{\prime }(y)+(\frac{A^{2}}{y}-%
\frac{B^{2}}{1-y}+C)\Re (y)=0.  \tag{A1}  \label{A1N}
\end{equation}

We can also redefine the function $\Re (y)$\ as \cite{Fernando}

\begin{equation}
\Re (y)=y^{\alpha }(1-y)^{\beta }F(y).  \tag{A2}  \label{A2N}
\end{equation}

Thus, the radial equation (A1) becomes

\begin{equation}
y(1-y)F^{\prime \prime }(y)+\left[ 1+2\alpha -2y(\alpha +\beta +1)\right]
F^{\prime }(y)+\left( \frac{\widetilde{A}}{y}-\frac{\widetilde{B}}{1-y}+%
\widetilde{C}\right) F(y)=0,  \tag{A3}  \label{A3N}
\end{equation}

where

\begin{equation}
\widetilde{A}=A^{2}+\alpha ^{2},  \tag{A4}  \label{A4N}
\end{equation}

\begin{equation}
\widetilde{B}=B^{2}-\beta ^{2},  \tag{A5}  \label{A5N}
\end{equation}

\begin{equation}
\widetilde{C}=C-(\alpha +\beta )^{2}-(\alpha +\beta ).  \tag{A6}  \label{A6N}
\end{equation}

If we set

\begin{equation}
\alpha =iA,\text{ \ \ }\rightarrow \text{ \ \ }\widetilde{A}=0,  \tag{A7}
\label{A7N}
\end{equation}

\begin{equation}
\beta =-B,\text{ \ \ }\rightarrow \text{ \ \ }\widetilde{B}=0,  \tag{A8}
\label{A8N}
\end{equation}

Eq. (A3) is transformed to the following second order differential equation

\begin{equation}
y(1-y)F^{\prime \prime }(y)+[1+2iA+2(B-1-iA)y]F^{\prime }(y)+\left[
(A+iB)^{2}+B+C-iA\right] F(y)=0.  \tag{A9}  \label{A9N}
\end{equation}

The above equation resembles the hypergeometric differential equation which
is of the form \cite{Abram}

\begin{equation}
y(1-y)F^{\prime \prime }(y)+[c_{n}-(1+a_{n}+b_{n})y]F^{\prime
}(y)-a_{n}b_{n}F(y)=0.  \tag{A10}  \label{A10N}
\end{equation}

Eq. (A10) has two independent solutions, which can expressed around three
regular singular points: $y=0$, $y=1$, and $y=\infty $ as \cite{Abram}

\begin{equation}
F(y)=C_{1}F\left( a_{n},b_{n};c_{n};y\right) +C_{2}y^{1-c_{n}}F\left(
a_{n}-c_{n}+1,b_{n}-c_{n}+1;2-c_{n};y\right)  \tag{A11}  \label{A11N}
\end{equation}

By comparing the coefficients of Eqs. (A9) and (A10), one can obtain the
following identities:

\begin{equation}
c_{n}=1+2iA,  \tag{A12}  \label{A12N}
\end{equation}

\begin{equation}
1+a_{n}+b_{n}=-2(B-1-iA),  \tag{A13}  \label{A13N}
\end{equation}

\begin{equation}
a_{n}b_{n}=-(A+iB)^{2}-B-C+iA.  \tag{A14}  \label{A14N}
\end{equation}

Solving system of Eqs. (A13) and (A14), we get

\begin{equation}
a_{n}=\frac{1}{2}-B+iA+\frac{\sqrt{4C+1}}{2},  \tag{A15}  \label{A15N}
\end{equation}

\begin{equation}
b_{n}=\frac{1}{2}-B+iA-\frac{\sqrt{4C+1}}{2}.  \tag{A16}  \label{A16N}
\end{equation}

Using Eqs. (A2), (A11), (A12), (A15), and (A16), one obtains the general
solution of Eq. (28) as Eq. (35).

\end{document}